\documentclass{PoS}
\usepackage{graphicx}
\newcommand{\be}{\begin{equation}}
\newcommand{\ee}{\end{equation}}
\newcommand{\bea}{\begin{eqnarray}}
\newcommand{\eea}{\end{eqnarray}}
\newcommand{\<}{\langle}
\renewcommand{\>}{\rangle}
\renewcommand{\[}{\langle\!\langle}
\renewcommand{\]}{\rangle\!\rangle}

\PoS{PoS(LAT2005)216}

\title{%
\begin{picture}(0,0)(0,0)%
\put(0,75){\makebox(0,0)[l]{\textnormal{\normalsize DESY 05-192}}}%
\put(0,60){\makebox(0,0)[l]{\textnormal{\normalsize RCNP-Th05029}}}%
\end{picture}%
Determination of the spin-dependent potentials with
the multi-level algorithm}

\ShortTitle{Determination of the spin-dependent potentials 
with the multi-level algorithm}

\author{\speaker{Miho Koma}\\
Deutsches Elektronen-Synchrotron DESY, 22603 Hamburg, Germany\\
Research Center for Nuclear Physics (RCNP), 
Osaka University, Ibaraki 567-0047, Japan\\
E-mail: ~\email{miho.koma@desy.de}}

\author{Yoshiaki Koma, Hartmut Wittig\\
Deutsches Elektronen-Synchrotron DESY, 22603 Hamburg, Germany\\
E-mail: \email{yoshiaki.koma@desy.de},~\email{hartmut.wittig@desy.de}}

\abstract{The spin-dependent corrections to the static interquark 
potential are relevant to describing the fine and 
hyper-fine splittings of the heavy quarkonium spectra.
We investigate these corrections in SU(3) lattice gauge theory 
with the Polyakov loop correlation function
as the quark source by applying the multi-level algorithm.
We observe remarkably clean signals for the
spin-dependent potentials up to intermediate distances.
}

\FullConference{XXIIIrd International Symposium on Lattice Field 
Theory\\
25-30 July 2005\\
Trinity College, Dublin, Ireland}

\begin{document}

\section{Introduction}

The spin-dependent (spin-orbit, spin-spin) 
interquark potentials are
relevant to describing the fine and hyper-fine splittings
of the heavy quarkonium  spectra and thus it is interesting to 
determine their behavior directly from QCD.
Eichten and Feinberg~\cite{Eichten:1979pu}
derived in this context the general form of the 
potential including the spin-dependent
corrections up to $O(1/m^{2})$,
\bea
V(r) &=& V_{0}(r) 
+
\left (
   \frac{\vec{s}_{1}\cdot \vec{l}_{1}}{2m_{1}^{2}}
- \frac{\vec{s}_{2}\cdot \vec{l}_{2}}{2m_{2}^{2}}
\right )
\left (  \frac{V_{0}'(r)}{r} +2 \frac{V_{1}'(r)}{r} \right )
+
\left (
  \frac{\vec{s}_{2}\cdot \vec{l}_{1}}{2 m_{1}m_{2}}
+\frac{\vec{s}_{1}\cdot \vec{l}_{2}}{2m_{1}m_{2}}  
\right )
\frac{V_{2}'(r)}{r}
\nonumber\\*
&&
+
\frac{1}{m_{1}m_{2}}
\left ( 
\frac{(\vec{s}_{1}\cdot \vec{r})( \vec{s}_{2}\cdot \vec{r})}
{r^{2}} -\frac{\vec{s}_{1}\cdot \vec{s}_{2}}{3} \right ) V_{3}(r)
+
\frac{\vec{s}_{1}\cdot \vec{s}_{2}}{3m_{1}m_{2}}V_{4}(r)\; ,
\eea
where the locations of quark and antiquark are
$\vec{r}_{1}$, $\vec{r}_{2}$ ($r \equiv |\vec{r}_{1}-\vec{r}_{2}|$).
$m_{1}$, $m_{2}$ ($=m$) denote their masses,
$\vec{s}_{1}$, $\vec{s}_{2}$ the spins, 
$\vec{l}_{1} = -\vec{l}_{2}=\vec{l}$
the orbital angular momenta.
$V_{0}(r)$ is the spin-independent static potential and
$V_{i}(r)$ ($i=1,\ldots,4$) the spin-dependent 
potentials, which are expressed in terms of 
the correlation function of two field
strength operators 
attached to the quark and antiquark, respectively.
The prime denotes the derivative with respect to~$r$.

\par
The determination of these potentials through 
lattice Monte Carlo simulations goes back
to the 1980s~\cite{deForcrand:1985zc,Michael:1985rh,%
Campostrini:1986ki,Huntley:1986de}.
The latest investigations are found in refs.~\cite{Bali:1996cj,Bali:1997am}.
The qualitative (quantitative to some extent) 
findings which seem to be established
are that while the spin-orbit potential $V_{1}(r)$ contains
the long-ranged nonperturbative component, 
all other potentials are only relevant to the short range
as explained by one-gluon exchange interaction.
However, the observed spin-dependent potentials
from even the latest studies~\cite{Bali:1996cj,Bali:1997am}
suffer from large numerical errors, which obscure
the behaviors already at intermediate distances.
For the phenomenological use of these potentials, 
it is clearly important to determine the form
of the potentials as accurately  as possible.

\par
For this purpose, we employ the multi-level 
algorithm~\cite{Luscher:2001up}
with a certain modification as applied to the measurement 
of the electric-flux profile between static 
charges~\cite{Koma:2003gi}.
The problem is quite similar to this, since we need to measure
the correlation function between the quark source and the
field strength operator.
This algorithm also allows us to 
use the Polyakov loop correlation function (PLCF:
a pair of Polyakov loops $P$ separated by a distance $r$) as 
the quark source instead of the Wilson loop.
We use  the field strength operator defined
by $F_{\mu\nu}=(U_{\mu\nu}-U_{\mu\nu}^{\dagger})/2i$,
where $U_{\mu\nu}$ is the plaquette variable.
The electric and magnetic fields are then
$E_{i}=F_{4i}$ and $B_{k}=F_{ij}$.
Noting $\[ F_{\mu\nu} F_{\rho\sigma}\] \equiv 
\< F_{\mu\nu} F_{\rho\sigma} \>_{P^{\dagger}P}
/ \< P^{\dagger}P\>$,
where $\< F_{\mu\nu} F_{\rho\sigma} \>_{P^{\dagger}P}$ 
is the two field strength correlator with the PLCF background,
the spin-dependent potentials with the PLCF, for $\vec{r}=(r,0,0)$,
are expressed as
\bea
&&
V_{1}'(r)= 2 \int_{0}^{\infty} d\tau \; \tau
\[ B_{y}(\vec{r}, 0) E_{z}(\vec{r},\tau) \]  \; , 
\label{eqn:pot1}\\
&&
V_{2}'(r) = 2\int_{0}^{\infty} d \tau \; \tau
\[   B_{y}(\vec{0},0) E_{z}(\vec{r},\tau) \]  \; ,  
\label{eqn:pot2}\\
&&
V_{3}(r) = 2\int_{0}^{\infty} d\tau \;
[\[ B_{x}(\vec{0},0)B_{x}(\vec{r},\tau) \]
-\[ B_{y}(\vec{0},0)B_{y}(\vec{r},\tau) \]  ] \; , 
\label{eqn:pot3}\\
&&
V_{4}(r) = 2\int_{0}^{\infty} d\tau \;
[\[ B_{x}(\vec{0},0)B_{x}(\vec{r},\tau) \]
+ 2 \[ B_{y}(\vec{0},0)B_{y}(\vec{r},\tau) \]  ] \; .
\label{eqn:pot4}
\eea
It is expected from these expressions that 
in contrast to the use of the Wilson loop as commonly 
applied in previous works,
the PLCF helps to reduce a systematic error associated with 
the limiting procedure in the integral, $\tau \to \infty$.
At least we will have data up to $\tau = T/2$, where $T$ is the temporal
size of the lattice volume.

\section{Numerical procedures} 

\begin{figure}[t]
\centering\includegraphics[width=6cm]{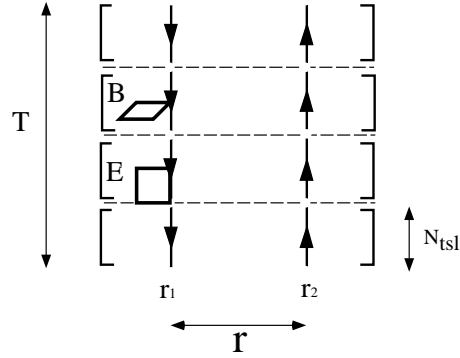}
\caption{How to construct 
$\<  B_{y}(r,0) E_{z}(r,\tau) \>_{P^{\dagger} P}$.
$[\cdots]$ denotes the sublattice average.
Other correlation functions are constructed in a similar way.} 
\label{fig:fig1}
\end{figure}

We describe the procedure how to compute the field strength 
correlator with the 
PLCF background using the multi-level algorithm 
(here we restrict to the lowest level).
The standard Wilson action is most preferable
for the multi-level algorithm because its action 
density is locally defined.
Thus we shall use this action in our simulation.
Periodic boundary conditions are imposed 
in all directions.
The essence of the multi-level algorithm is to 
construct the desired correlation function
from the ``sublattice average'' of its components.
In our case the corresponding parts are
the two-link correlator and the 
field-strength-inserted two-link correlator.
For a schematic understanding, see Fig.~\ref{fig:fig1}, which 
illustrates the computation of the correlation function,
$\<B_{y}E_{z}\>$, for $V_{1}'$.

\par
The sublattice is defined by dividing 
the lattice volume into several layers along the time direction and
thus a sublattice consists of a certain number of time slices.
We then take averages of the components of the correlation function
at each sublattice by updating the gauge field (with a mixture
of HB/OR), while the space-like links on the boundary
between sublattices remain intact during the update.
We repeat the sublattice update until we obtain stable 
signals for the components.
Then, we multiply these averaged components in a 
desired way and complete the correlation function.
This is how the correlation function is constructed
from ``one'' configuration.
We then update the whole links without specifying any
layers to obtain another independent gauge configuration
and start the above sublattice averaging for the next
configuration.

\par
In order to benefit from this algorithm, we need to optimize
the number of time slices in each sublattice $N_{\rm tsl}$,
and the number of the internal update $N_{\rm iupd}$ for 
sublattice averaging.
They depend on the coupling $\beta$
and on the distances to be investigated.
In principle, these parameters can be determined by 
looking at the behavior of the correlation function as a 
function of~$N_{\rm iupd}$ for several~$N_{\rm tsl}$.
An empirical observation shows that 
$a N_{\rm tsl} = 0.3 - 0.4$~fm is optimal in order 
to suppress the fluctuation
of the correlation function among configurations.

\section{Numerical results}

\begin{figure}[t]
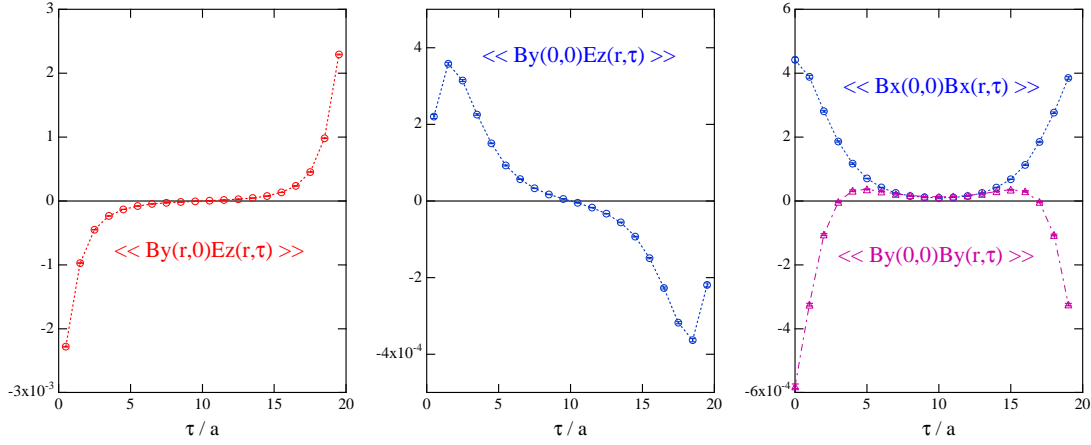

 \centering
 \includegraphics[width=4.8cm]{cor1.EPSF}
 \includegraphics[width=4.8cm]{cor2.EPSF}
 \includegraphics[width=4.8cm]{cor34.EPSF}
 \caption{Correlation functions as a function of $\tau$ 
at $r/a = 4$ on the $20^{4}$ lattice} 
\label{fig:fig2}
\end{figure}

We present the result obtained at $\beta=6.0$ 
($a \approx 0.09$ fm)  on the $20^{4}$ lattice, 
where the ranges of the measured
distances between static charges are $R = 2 - 7$.
At $\beta=6.0$, we found that $N_{\rm tsl}=4$ 
is the optimal choice.
We then chose $N_{\rm iupd}=7000$ 
to be able to see the signal at least up to $r/a=7$.
The number of configuration is $N_{\rm conf}=76$.
One Monte Carlo update consists of 1~HB/5~OR.
In Fig.~\ref{fig:fig2}, we show the typical behavior of 
the correlation functions as a function of $\tau$ at $r/a=4$.
We observe clean signals for the whole range of $\tau$.
We also obtained similar clean data for other distances.

\par
Once the correlation functions are obtained, 
our next task is to perform the integration 
in Eqs.~(\ref{eqn:pot1})$-$(\ref{eqn:pot4}) to obtain 
the potentials.
Since the integration range of $\tau$ is limited at most to $T/2$,
we need an extrapolation to extract the value 
corresponding to $\tau \to \infty$.
This procedure is in fact the potential source of 
the systematic error and needs careful analysis,
in particular, 
when the statistical errors are significantly small
as shown in Fig.~\ref{fig:fig2}.

\par
Currently we applied the following analysis.
Since the correlation functions were reasonably 
smooth, we firstly performed the cubic spline interpolation 
of the integrand and secondly  evaluated 
the integral analytically in the range $\tau \in [0,\tau_{\rm max}]$,
where $\tau_{\rm max} = 1,2,\ldots,T/2$.
Then, we fitted this result
with a function which has an asymptotic constant value
at $\tau_{\rm max} \to \infty$, like
$c+c' \exp (-c'' \tau_{\rm max})$ (exponential type)
or $c/(1+(c'/\tau_{\rm max})^{c''})$ (Hill type).
The validity of the fit and the choice of the fitting function
were monitored by looking at the
the minimum of $\chi^{2}$ defined
with the covariance matrix so as to take into account
the correlation among different $\tau$'s.
The errors are evaluated from the distribution of the jackknife
samples of the fitting parameters.

\par
We found that this method 
at least works well to extract values at $\tau_{\rm max} \to \infty$
 for $V_{1}'$, $V_{2}'$ and $V_{3}$.
The results are shown in Figs.~\ref{fig:fig3} and~\ref{fig:fig4} (left).
For $V_{4}$, however, we found that this method 
needs to be modified especially at intermediate distances, 
because we observed a peculiar 
finite $\tau$ effect due to the symmetric behaviors
of $\[ B_{x}B_{x}\]$ and $\[ B_{y}B_{y}\]$
at $\tau=T/2=10$.
Thus, we just plot the integration result at $\tau_{\rm max}=9$
in Fig.~\ref{fig:fig4} (right).
Systematic effects of the extrapolation
as well as finite volume effects 
will be investigated in future work.

\par
The qualitative behaviors of these potentials are that 
the spin-orbit potential $V_{1}'$ contains the long-ranged 
nonperturbative component ($V_{1}'(r)$ behaves as a constant 
at large $r$), while $V_{3}$ and~$V_{4}$ seem to be relevant  
at short distances.
These findings are in agreement with previous works.
However, the statistical errors are significantly reduced.
It is interesting to find that $V_{2}'$ is not restricted to the 
short range, rather it has a finite tail up to 
intermediate distances.

\begin{figure}[t]
\centering
\includegraphics[width=7cm]{pot1.EPSF}
\includegraphics[width=7cm]{pot2.EPSF}
\caption{The spin-orbit potentials $V_{1}'$ (left) and $V_{2}'$ (right).} 
\label{fig:fig3}
\includegraphics[width=7cm]{pot3.EPSF}
\includegraphics[width=7cm]{pot4.EPSF}
\caption{The spin-spin potentials $V_{3}$ (left) and $V_{4}$ (right).} 
\label{fig:fig4}
\end{figure}

\section{Summary and outlook}

We have measured the spin-dependent potentials 
in SU(3) lattice gauge theory 
with the Polyakov loop correlation function (PLCF)
by applying the multi-level algorithm.
The method presented here is promising to carry out 
further systematic investigations, 
such as the computation of the renormalization factor
of the field strength operators, $Z_{B}$ and $Z_{E}$,
as well as the scaling study,
which are both necessary for the discussion of the fine/hyper-fine 
structure of the heavy quarkonium spectra.

\par
The preliminary studies of the renormalization factors 
defined {\em a la} Huntley and Michael~\cite{Huntley:1986de},
but using the PLCF, show the similar values
as in ref.~\cite{Bali:1997am}.
We will report these issues in our forthcoming publication.
It is also interesting to examine the Gromes
relation~\cite{Gromes:1984ma}, $V_{0}'=V_{2}'-V_{1}'$, with 
high precision.

\par
Finally we note that this method is also applicable
to measuring the momentum-dependent potentials
up to $O(1/m^{2})$~\cite{Barchielli:1986zs,Pineda:2000sz} 
from the PLCF, since the correlation functions to be measured
are quite similar to that for 
the spin-dependent potentials.
This will help to refine the data reported in 
ref.~\cite{Bali:1997am}, which is in progress.

\section*{Acknowledgments}

We thank R.~Sommer and A.~Pineda for useful discussions.
The main calculation has been performed on the NEC SX5 
at Research Center for Nuclear Physics (RCNP), 
Osaka University, Japan.


\begin{thebibliography}{99}

\bibitem{Eichten:1979pu}
E.~Eichten and F.~Feinberg,
{\em Spin dependent forces in heavy quark systems},
Phys. Rev. Lett. {\bf 43} (1979) 1205.

\bibitem{deForcrand:1985zc}
Ph.~de~Forcrand and J.D. Stack,
{\em Spin dependent potentials in SU(3) lattice gauge theory},
Phys. Rev. Lett. {\bf 55} (1985) 1254.

\bibitem{Michael:1985rh}
C.~Michael,
{\em The long range spin orbit potential},
Phys. Rev. Lett. {\bf 56} (1986) 1219.

\bibitem{Campostrini:1986ki}
M.~Campostrini, K.~Moriarty, and C.~Rebbi,
{\em Monte carlo calculation of the spin dependent potentials for heavy quark
  spectroscopy},
Phys. Rev. Lett. {\bf 57} (1986) 44.

\bibitem{Huntley:1986de}
A.~Huntley and C.~Michael,
{\em Spin-spin and spin-orbit potentials from lattice gauge theory},
Nucl. Phys. {\bf B286} (1987) 211.

\bibitem{Bali:1996cj}
G.S. Bali, K.~Schilling, and A.~Wachter,
{\em Ab initio calculation of relativistic corrections to the static interquark
  potential. I: SU(2) gauge theory},
Phys. Rev. {\bf D55} (1997) 5309 
[{\tt hep-lat/9611025}].

\bibitem{Bali:1997am}
G.S. Bali, K.~Schilling, and A.~Wachter,
{\em Complete {$O(v^2)$} corrections to the static interquark potential from
  SU(3) gauge theory},
Phys. Rev. {\bf D56} (1997) 
2566  [{\tt hep-lat/9703019}]. 
    
\bibitem{Luscher:2001up}
M.~L{\"u}scher and P.~Weisz,
{\em Locality and exponential error reduction in numerical lattice gauge
  theory},
JHEP {\bf 09} (2001) 010 
[{\tt hep-lat/0108014}].

\bibitem{Koma:2003gi}
Y.~Koma, M.~Koma, and P.~Majumdar,
{\em Static potential, force, and flux-tube profile in 4D compact U(1) lattice
  gauge theory with the multi-level algorithm},
Nucl. Phys. {\bf B692} (2004) 209
[{\tt hep-lat/0311016}].

\bibitem{Gromes:1984ma}
D.~Gromes,
{\em Spin dependent potentials in QCD and the correct long range spin orbit
  term},
Z. Phys. {\bf C26} (1984) 401.

\bibitem{Barchielli:1986zs}
A.~Barchielli, E.~Montaldi, and G.M. Prosperi,
{\em On a systematic derivation of the quark-antiquark potential},
Nucl. Phys. {\bf B296} (1988) 625, 
{Erratum-{\em ibid}} {\bf B303} (1988) 752.
  .
\bibitem{Pineda:2000sz}
A.~Pineda and A.~Vairo,
{\em The QCD potential at {$O(1/m^2)$}: Complete spin-dependent and
  spin-independent result},
Phys. Rev. {\bf D63} (2001) 054007 
[{\tt hep-ph/0009145}], {Erratum-{\em ibid}} {\bf
  D64} (2001) 039902.

  
\end{thebibliography}
\end{document}